\documentclass{iaus}
\usepackage{graphicx}

\title[A Survey of M Stars in the Field of View of \emph{Kepler}] 
{A Survey of M Stars in the Field of View of {\emph{Kepler}} Space Telescope}

\author[Oshagh, Haghighipour, \& Santos]   
{Mahmoudreza Oshagh$^1$, Nader Haghighipour$^2$, and Nuno C.Santos$^1$}

\affiliation{$^1$Centro de Astrof\'\i sica, Faculdade de Ci\^encias,
Universidade do Porto, Rua das Estrelas, 4150--762 Porto,
Portugal \\ email: {\tt moshagh@astro.up.pt} \\[\affilskip]
$^2$Institute for Astronomy and NASA Astrobiology Institute, University of Hawaii-Manoa,
2680 Woodlawn Drive, Honolulu, HI 96822,USA}

\pubyear{2010}
\volume{276}  
\jname{The Astrophysics of Planetary Systems: Formation, Structure, and
Dynamical Evolution}
\editors{A. Sozzetti, M. Lattanzi \&\ A.P. Boss , eds.}
\begin{document}

\maketitle

\begin{abstract}
M dwarfs constitute more than $70\% $ of the stars in the solar neighborhood.
They are cooler and smaller than Sun-like stars and have less-massive disks which
suggests that planets around these stars are more likely to be Neptune-size or
smaller. The transit depths and transit times of planets
around M stars are large and well-matched to the \emph{Kepler} temporal resolution.
As a result, M stars have been of particular interest for searching for planets in both radial
velocity and transit photometry surveys. We have recently started a project on searching
for possible planet-hosting M stars in the publicly available
data from \emph{Kepler} space telescope.
We have used four criteria, namely, the
magnitude, proper motion, $H$-${K_s}$ and $J$-$H$ colors, and searched for M stars in
Q0 and Q1 data sets. We have been able to find 108 M stars among which 54 had not been
previously identified among \emph{Kepler}'s targets. We discuss the details of our
selection process and present the results.

\end{abstract}

\firstsection 
              
\section{Introduction}

The \emph{Kepler} space telescope is monitoring more than 150,000 stars in a 105 square degree 
field of view around the
constellations Cygnus and Lyra. The data from this telescope provide a great opportunity for
identifying potential terrestrial planets around variety of stars. Among these stars, M dwarfs 
are ideal targets for searching for
terrestrial/habitable planets. These stars have the greatest reflex acceleration due to an
orbiting planet, their low surface temperatures place their habitable zones
at close distances, and their light curves show large decrease when they are
transited by a planet. The transit depths and transit times of planets around M stars are also large
and well-matched to the \emph{Kepler} temporal resolution. We have surveyed \emph{Kepler}'s
publicly available data from quarters Q0 and Q1, and identified more than a hundred
M stars. In this paper, we discuss our methodology and present our selection process.

\section{Selection Criteria}
\cite{TraubCutri08} were the first to make an attempt to identify M stars in
\emph{Kepler}'s field of view. They considered stars with $K_{s}$  smaller than 13.5 mag,
and proposed a selection process based on a criterion on $H$-$K_{s}$ vs. $J$-$H$
colors. Since M stars
are relatively faint and abundant, \cite{TraubCutri08} suggested that any detected M star
is likely to be nearby and therefore likely to have a relatively large proper motion.
Using these criteria, these authors estimated that close to 1600 M stars should exist
in \emph{Kepler}'s field of view. A recent article by \cite{Batalha_etal10} suggests 
that the actual number of M stars
in  \emph{Kepler}'s field of view  may be twice as large.
Using selection criteria based on a star's surface gravity, effective
temperature, and inferred stellar radius, these authors were able to show that
approximately 3000 M stars are present in the field of view of \emph{Kepler}.

We used the following criteria to look for M stars in \emph{Kepler}'s
publicly available data of quarters Q0 and Q1. These criteria are somewhat similar
to those of \cite{TraubCutri08} with the exception that we expanded the requirements
on the magnitudes of stars using the ranges suggested by \cite{Leggett92}:

\vskip 5pt

\begin{itemize}
\item $K$ magnitude equal to or smaller than 15 mag,
\item $J$-$H$ color in the range of 0.42 to 0.78,
\item $H$-$K$ color in the range of 0.12 to 0.50,
\item Proper motion bigger than 0.1.

\end {itemize}

\vskip 5pt

\noindent
Our search resulted in identifying 108 M stars. Among these stars, 27 are
in both the Q0 and Q1 data sets. When comparing with the sample of M stars presented by
\cite{Batalha_etal10}, we discovered that there are no data on the effective temperature
and surface gravity of 54 of our M stars. As a result, these 54 M stars were not
identified by \cite{Batalha_etal10}. One example of these M stars is LHS 6343, a member
of a M star binary system with a transiting brown dwarf (\cite{Johnson10}).
Table 1 shows 46 of these 54 M stars that have $\log g>3.5$. These stars have been
binned by magnitude and effective temperature.

\vskip 10pt
\noindent
This work has been supported by the European Research Council/European Community
under the FP7 through a Starting Grant, as well as in the form of grant reference 
PTDC/CTEAST/098528/2008 funded by Funda\c{c}$\tilde{a}$o para a Ci\^encia e a Tecnologia
(FCT), Portugal. NCS would further like to thank the support from FCT through a
Ci\^encia , 2007 contract funded by FCT/MCTES (Portugal) and POPH/FSE (EC).
NH acknowledges support from NASA Astrobiology Institute (NAI) under Cooperative Agreement
NNA04CC08A at the Institute for Astronomy, University of Hawaii, NAI central, and NASA EXOB
grant NNX09AN05G.

\begin{table}[]
\caption{Star Counts as a Function of Effective Temperature and Magnitude}
\centering 
\begin{tabular}{c c c c c c c} 
\hline
Mag & & $T_{\rm eff}$ & & $T_{\rm eff}$  & &  Total \\
[0.5ex]
& & $(4500-5500)$ & & $(3500-4500)$ & & \\
\hline 
$6<K  <7$  & & 0 & & 0 & & 0 \\ 
$7<K  <8$  & & 0 & & 0 & & 0 \\
$8<K  <9$  & & 0 & & 5 & & 5 \\
$9<K  <10$ & & 1 & & 3 & & 4 \\
$10<K <11$ & & 0 & & 9 & & 9 \\
$11<K <12$ & & 0 & & 13 & & 13 \\
$12<K <13$ & & 1 & & 13 & & 14 \\
$13<K <14$ & & 0 & & 1 & & 1 \\
$14<K <15$ & & 0 & & 0 & & 0 \\
[1ex] 
\hline
\end{tabular}
\label{table:nonlin} 
\end{table}


\begin{thebibliography}{}





\bibitem[Batalha et al. (2010)]{Batalha_etal10}
Batalha, N. M., et al. 2010, \textit{ApJ}, 713, L109

\bibitem[Johnson et al. (2010)]{Johnson10}
Johnson, J. A., et al. 2010, (astroph:1008.4141v3)

\bibitem[Leggett (1992)]{Leggett92}
{Leggett, S. K.} 1992, \textit{ApJS}, 82, 351

\bibitem[Traub \& Cutri (2008)]{TraubCutri08}
{Traub, W., \& Cutri, R.} 2008, in: Extreme Solar Systems, Eds. D. Fischer, F. A. Rasio,
S. E., Thorsett, and A. Wolszczan, ASP Conference Series, 398, 475








\end{thebibliography}
\end{document}